\documentstyle[11pt,epsf]{article}

\newcommand{\be}{\begin{equation}}
\newcommand{\ee}{\end{equation}}
\newcommand{\bea}{\begin{eqnarray}}
\newcommand{\eea}{\end{eqnarray}}

\begin{document}

\title{{\Large \bf SYMPLECTIC VS PSEUDO-ORTHOGONAL\\
STRUCTURE OF SPACE-TIME}%
\thanks{Report presented  at XXIV International
Workshop on Fundamental Problems in HEP and Field Theory, 27-29 June
2001, Protvino, Russia; to appear in the proceedings.} 
}
\author{Yu.F.\ Pirogov \\[0.5ex]
{\it Institute for High Energy Physics,}\\
{\it Protvino, Moscow Region RU-142284, Russia}}
\date{}
\maketitle

\abstract{\noindent
The advantages to consider the ordinary space-time as
the symplectic  rather than pseudo-orthogonal one are indicated,
and the consequences of
extending this view to extra space/time dimensions are discussed.}

\section{Symplectic vs pseudo-orthogonal space-time}

The space-time, or the world  we live in is generally adopted to be
(locally) the
Minkowski one. Its structure group is the pseudo-ortho\-gonal group
$SO(1,3)$. To a space-time point there corresponds a real four-vector.

On the other hand, the spinor calculus in our space-time heavily
relies on the isomorphy of the noncompact groups $SO(1,3)\simeq
SL(2,C)/Z_2$, as well as that $SO(3)\simeq SU(2)/Z_2$ for their
maximal compact subgroups. In fact, the whole relativistic field
theory in four space-time dimensions can
equivalently be formulated in the framework of the complex unimodular
group $SL(2,C)$ alone. In a sense, it is even preferable. In this
approach, to a space-time point there corresponds a Hermitian
spin-tensor of the second-rank.

There is a choice for the ordinary space-time structure group: 
either the
pseudo-orthogo\-nal symmetry $SO(1,3)$, with vectors as defining
representation and spinors as a kind of artifact, or the complex
symplectic group $Sp(2,C)\simeq SL(2,C)$ with spinors as defining
representations and vectors as a secondary object. The two approaches
are mathematically equivalent. Nevertheless, the symplectic approach
seems physically more appropriate.

Then, in searching for the space-times with extra dimensions it is
natural to look for the extensions in the symplectic framework with
the structure group $Sp(2l,C)$, $l>1$ instead of $SO(1,d-1)$, $d>4$.
The symplectic series of the groups (contrary to $SL(l+1,C)$) is
peculiar quantum-mechanically
because it retains the invariant bilinear spinor product at any $l>1$.

Two alternative directions of the space-time extension can
schematically be
pictured  as follows: 
\begin{eqnarray}
d=4\qquad\hspace*{4ex} SO(1,3)& \simeq&{\ Sp(2,C)}\qquad\ l=1
\nonumber \\
\downarrow\ \ \ \ &&\ \ \ \ \ \downarrow  \nonumber \\
d>4\qquad SO(1,d-1)& \not\simeq&{\ Sp(2l,C)}\qquad l>1\,.  \nonumber
\end{eqnarray}
In the pseudo-orthogonal direction of extension, the local metric
properties
of the space-times, i.e., their dimensionalities and signatures, are
to be put in from the very beginning. In the symplectic direction,
these
properties are not to be considered as the primary ones but, instead,
they should emerge as a
manifestation of the adopted symplectic structure\footnote{For more
detail we refer to the literature: Yu.F.~Pirogov, IHEP 2001-19 (2001),
hep-ph/0104119.}.

\section{General symplectic framework}

\paragraph*{Arbitrary symplectic space-time: \protect\boldmath{l = 1,
2, \dots}}

Let $\psi _A$ and $\bar \psi {}^{\bar A}\equiv (\psi _A)^{*}$, as well
as their respective duals $\psi ^A$ and $\bar \psi {}_{\bar A}\equiv
(\psi ^A)^{*}$, with indices $A$, $\bar A=1,\dots ,2l$, are the spinor
representations of $Sp(2l,C)$. There exist the invariant second-rank
spin-tensors $\epsilon _{AB}=-\epsilon _{BA}$ and
$\epsilon^{AB}=-\epsilon
^{BA}$ such that $\epsilon _{AC} \epsilon ^{CB}=\delta _A{}^B$, with 
$\delta_A{}^B$ being the Kroneker symbol (and similarly for 
$\epsilon_{\bar A\bar B}\equiv (\epsilon ^{BA})^{*}$ and 
$\epsilon ^{\bar A\bar B}\equiv (\epsilon_{BA})^{*}$). Owing to these
tensors the spinor indices of the
upper and lower positions are pairwise equivalent ($\psi _A\equiv
\epsilon_{AB}\psi^B$ and $\bar \psi _{\bar A}\equiv 
\epsilon_{\bar A\bar B}\bar \psi^{\bar B}$), so that there are left
just two inequivalent spinor representations (generically, $\psi $ and
$\bar \psi $). These are spinors
of the first and the second kind, respectively.

Let us put in correspondence to an event point $P$ a
second-rank $2l\times 2l$ spin-tensor $X_A{}^{\bar B}(P)$, which is
Hermitian, i.e., fulfil the restriction 
\[
X_A{}^{\bar B}= (X_B{}^{\bar A})^*\equiv\bar X^{\bar B}{}_A\,, 
\]
or in other terms 
\[
X^{A\bar B}= (X_{B\bar A})^*\equiv \bar X^{\bar B A}. 
\]

The quadratic scalar product is defined as follows: 
\[
\label{eq:X2} (X,X)\equiv\mbox{tr\,}X\bar X= X_A{}^{\bar B}\bar
X_{\bar
B}{}^A= -X_{A\bar B}(X_{B\bar A})^*\,. 
\]
Here $(X,X)$ is real though not sign definite. Under arbitrary
transformation $S\in Sp(2l,C)$ one has in short notations: 
\begin{eqnarray}
X& \to &SXS^\dagger\,,  \nonumber \\
\bar X&\to & S^{\dagger -1}\bar X S^{-1}  \nonumber
\end{eqnarray}
and hence $(X,X)$ is invariant. At $l>1$, the quadratic invariant
above is
just the lowest order one in a series of independent invariants
$\mbox{tr\,}(X\bar X)^k$, $k=1,\dots,l$. The highest order one with
$k=l$ is equivalent to $\mbox{det\,}X$.

\vspace{1ex} \noindent
{\bf Definition:} the Hermitian spin-tensor set $\{X\}$ equipped with
the structure group $Sp(2l,C)$ and the interval between points $X_1$
and $X_2$ equal to $(X_1-X_2,X_1-X_2)$ constitutes the flat symplectic
space-time.
\vspace{1ex}

The noncompact transformations out of $Sp(2l,C)$ are counterparts
of the Lorentz boosts in the ordinary space-time $l=1$, while
transformations out of the compact subgroup $Sp(2l)=Sp(2l,C)\cap
SU(2l)$ correspond to rotations.
With account for translations $X_A{}^{\bar B}\to X_A{}^{\bar
B}+\Xi_A{}^{\bar B}$, where $\Xi_A{}^{\bar B}$ is an arbitrary
constant Hermitian spin-tensor, the whole theory in the flat
symplectic space-time should be covariant relative to the
inhomogeneous symplectic group $ISp(2l,C)$.

Restricting by the maximal compact subgroup $Sp(2l)$, the indices
of the first and the second kinds in the same position are
indistinguishable relative to their transformation properties. Hence,
under $Sp(2l)$ one can
reduce the event tensor $X_{A\bar B}$ into two irreducible parts,
symmetric and
antisymmetric ones: $X=X_{+}+X_{-}$, where 
$X_{\pm }=\pm (X_{\pm })^T$ have dimensionalities 
$d_{\pm }=l(2l\pm 1)$, respectively.
The scalar product decomposes as follows:
\[
(X,X)=\sum_{\pm }(\mp 1)(X_{\pm })_{A\bar B}[(X_{\pm })_{A\bar
B}]^{*}\,.
\]
Thus one of two pieces $X_{\pm }$ is the spatial part of coordinate
while the rest is the time part.

At $l>1$, one can further reduce the antisymmetric part $X_-$ of the
event spin-tensor into the trace relative to $\epsilon$ and a
traceless part: $X_-=X_-^{(0)}+X_-^{(1)}$. The trace $X_-^{(0)}$ is
rotationally invariant and hence represents the true time. 
In short: 
\[
\mbox{1-time} = \mbox{trace}\,. 
\]
The traceless part $X_-^{(1)}$ is uniquely associated with extra
times.  

Relative to the rotational subgroup, the whole extended space-time can
be decomposed  into three irreducible subspaces of the
dimensionalities $1$, $(l-1)(2l+1)$ and $l(2l+1)$, respectively. The
first two subspaces correspond to  time directions, while the third
subspace corresponds to the spatial ones.

Though any particular decomposition into $X_{\pm}$ is noncovariant and
depends on the boosts, the number of the positive and negative
components in $(X,X)$ is invariant under the whole $Sp(2l,C)$. In
other words, there emerges the invariant metric tensor of the
$d$-dimensional flat space-time: 
\[
\eta_d=(\,\underbrace{+1,\dots}_{d-}\,;\underbrace{- 1,
\dots}_{d_+}\,)\,. 
\]

Thus, at $l>1$ the structure group $Sp(2l,C)$, acting on the
Hermitian second-rank spin-tensors with $d=4l^2$ components, is a
subgroup of the
embedding pseudo-orthogonal group $SO(d_-,d_+)$, acting on the
pseudo-Euclidean space of the same dimensionality. What distinguishes
$Sp(2l,C)$ from $SO(d_-,d_+)$, is the total set of independent
invariants $\mbox{tr}(X\bar X)^k$, $k=1,\dots,l$. The isomorphy
between the groups is valid only at $l=1$, i.e., for the ordinary
space-time $d=4$ where there is just one invariant.

In the symplectic approach, neither the discrete set of
dimensionalities, $d=4l^2$, of the extended space-time, nor its
signature, nor the existence of
the rotationally invariant one-dimensional time subspace are
postulated from the beginning. These properties are the attributes of
the very symplectic structure.

In particular, the symplectic structure provides the simple rationale
for the four-dimension\-ality of the ordinary space-time, as well as
for its signature {\ ($+--\, -$)}. Namely, these properties just
reflect the existence of one antisymmetric and three symmetric 
$2\times 2$ Hermitian spin-tensors. In short: 
\[
(2\times 2)_{\rm H} = 1_{\rm A}\oplus 3_{\rm S} \,. 
\]
The set of the second-rank Hermitian tensors, in its turn, is the
lowest admissible real space to accommodate the symplectic structure.

On the other hand, right the one-dimensionality of time allows one to
put  the events in order at any fixed boosts and hence to insure the
causal description. Therefore, the causality might ultimately be
attributed to the underlying symplectic structure, too. At $l>1$, the
one-dimensional time and the extra times mix with each other via
boosts. Because of this influence of extra times, the causality is
expected not to fulfil for relativistic  events. 

\paragraph*{Gauge interactions} 

Let $D_A{}^{\bar B}\equiv\partial_A{}^{\bar B}+igG_A{}^{\bar B}$ be
the generic covariant derivative, where $\partial_A{}^{\bar B}\equiv
\partial/\partial X^A{}_{\bar B}$ is the ordinary derivative, $g$ is
the gauge coupling and the Hermitian spin-tensor $G_A{}^{\bar B}$ is
the gauge fields. One can introduce the gauge invariant strength
tensor 
\[
F_{\{A_1A_2\}}^{[\bar B_1\bar B_2]}= \partial_{\{A_1}^{[\bar B_1}
G_{A_2\}}^{\bar B_2]}+ ig G_{\{A_1}^{[\bar B_1} G_{A_2\}}^{\bar
B_2]}\,. 
\]

The total number of the real components in tensor $F$ precisely
coincides with the number of components of the antisymmetric
second-rank tensor $F_{[\mu\nu]}$, $\mu,\nu=0,1,\dots,4l^2-1$ in the
embedding pseudo-Euclidean space of the dimensions $d=4l^2$. But in
the symplectic case, tensor $F$ is reducible and splits into a trace
relative to $\epsilon$ and a traceless part, $F=F^{(0)}+F^{(1)}$.

For an unbroken gauge theory with fermions, the generic gauge, fermion
and mass terms of the Lagrangian ${\cal L}= {\cal L}_G+{\cal
L}_F+{\cal L}_M$ are, respectively, 
\begin{eqnarray}  \label{eq:L}
{\cal L}_G&=&\sum_{s=0,1}(c_s +i\theta_s)\, F^{(s)}F^{(s)}+{\rm h.c.}
\,, 
\nonumber \\[-0.5ex]
{\cal L}_F&=&\frac{i}{2}\sum_{\pm}(\psi ^{\pm})^\dagger\!
\stackrel{\leftrightarrow}{D }\psi^{\pm}\,,  \nonumber \\
{\cal L}_M&=&\psi^+m_0\,\psi^-
+\sum_{\pm}\psi^{\pm}m_{\pm}\psi^{\pm}+{\rm h.c.}\,,  \nonumber
\end{eqnarray}
where $\psi^{\pm}$ are pairs of the charged conjugate fermions.

The Lagrangian results in the following generalization of the Dirac
equation 
\[
\label{eq:dir} iD^C\!{}_{\bar B}\psi^{\pm}_C= m_0^\dagger
\overline\psi{}^{\pm}_{\bar B} +\sum_{\pm} m_\pm^\dagger\, \overline
\psi{}^{\mp}_{\bar B} 
\]
and the pair of Maxwell equations ($c_0\equiv 1$ and $c_1=\theta_1=0$
for simplicity) 
\begin{eqnarray}  \label{eq:max}
(1 +i\theta_0)D^{C\bar B}F^{(0)}{}_{\{C A \}}-\mbox{h.c.}&=&0\,,
\nonumber
\\
(1 +i\theta_0)D^{C\bar B}F^{(0)}{}_{\{C A\}}+\mbox{h.c.} &=&2 g
J_{A}{}^{\bar B}\,,  \nonumber
\end{eqnarray}
with $J_{A}{}^{\bar B}\equiv\sum_{\pm} (\pm 1) 
\psi^{\pm}_{A}({\psi}{}^\pm_{B})^\dagger$ being the fermion Hermitian
current.

Tensors $F^{(s)}$, $s=1,2$ are non-Hermitian, but under restriction by
the maximal compact subgroup $Sp(2l)$ they split into a pair of the
Hermitian
ones: $F^{(s)}=E^{(s)}+iB^{(s)}$. There is the duality transformation
$F^{(s)}\to \tilde F^{(s)}\equiv iF^{(s)}$, so that $\tilde
E^{(s)}=-B^{(s)}$
and $\tilde B^{(s)}=E^{(s)}$. Though the splitting into $E^{(s)}$ and
$B^{(s)}$ is noncovariant with respect to the whole $Sp(2l,C)$, the
duality transformation is covariant.

Tensors $E^{(s)}$ and $B^{(s)}$ are the counterparts of the ordinary
electric and magnetic strengths, and $\theta_0$, $\theta_1$ are the
$T$-violating $\theta$-parameters. Due to electric-magnetic
duality, the electric and magnetic strengths stay  in the framework of
symplectic extension on equal footing. This is to be contrasted with
the pseudo-orthogonal extension where these strengths have unequal
number of components at~$d\neq 4$.

\paragraph*{Gravity}

The considerations above refer to the flat extended space-time or,
otherwise, to the inertial local frames. To go beyond, one can
introduce the Hermitian local fielbeins $e_M{}_A{}^{\bar B}(X)$, with
$M=0,1,\dots, 4l^2-1$
being the world vector index, and the real world coordinates
$x_M\equiv \mbox{tr}X\bar e_M$. Now, one can equip space-time with the
pseudo-Riemannian structure, i.e., the real symmetric metrics
$g_{MN}(x)= \mbox{tr\,}e_M\bar e_N$. Introducing the generally
covariant derivative $\nabla\!_M(e)$ one can adapt the field theory to
the $d=4l^2$ dimensional curved space-time.

One can also supplement gauge equations by the generalized
Hilbert-Einstein gravity equations. But now the group of equivalence
of the local fielbeins (structure group) is just the symplectic group
$Sp(2l,C)$ rather than the whole pseudo-orthogonal group
$SO(d_-,d_+)$. This permits more independent
components in the local symplectic fielbeins compared to the metrics.
The curvature tensor in the symplectic case splits additionally into
irreducible parts which can enter the gravity Lagrangian with the
independent coefficients. Hence, the symplectic gravity is in general
not equivalent to the metric one. 

The reason for this may be as follows. In the symplectic approach, the
vectors and space-time in its present meaning are to be understood
not as the fundamental entities. Therefore, gravity treated
as a generally covariant theory of the space-time distortions have to
be just an effective theory. The latter admits a lot of free
parameters, the choice of which should eventually be clarified by an
underlying theory.

\section{Next-to-ordinary symplectic space-time: 
\protect\boldmath l = 2}

\paragraph*{Coordinate space kinematics}

In this case there takes place the isomorphy $SO(5,C)$ $\simeq
Sp(4,C)/Z_2$. Cases $l=1, 2$ are the only ones when the structure of
the symplectic group gets simplified in terms of the complex
orthogonal groups.

One can introduce the set of Clifford matrices $(\Sigma_I)_A{}^{\bar
B}$, $I=1,\dots,5$. Re\-lative to the maximal compact subgroup
$SO(5)$, they can
be chosen both Hermitian $(\Sigma_I)_A{}^{\bar B}= [(\Sigma_I)_{\bar
B}{}^{A}]^*$ and antisymmetric $(\Sigma_I)_{A\bar B}=-(\Sigma_I)_{\bar
B A}$, similar to $(\Sigma_0)_{AB}=\epsilon_{AB}$. One can require
that $\Sigma_I$ are traceless. Thus under $SO(5)$, six matrices
$\Sigma_0$, $\Sigma_I$ provide the complete independent set for the
antisymmetric matrices in the four-dimensional spinor space.

After introducing matrices $\Sigma_{IJ}=-i/2
(\Sigma_I\overline\Sigma_J-\Sigma_J\overline \Sigma_I)$, with $\bar
\Sigma\equiv- \epsilon\Sigma^*\epsilon$\,, one gets the (anti)symmetry
conditions for them: $\Sigma_{IJ}=-\Sigma_{JI}$ and
$(\Sigma_{IJ})_{AB}=(\Sigma_{IJ})_{B A}$. Therefore, ten matrices
$\Sigma_{IJ}$ make up the
complete set for the symmetric matrices in the spinor space. The
matrices ($\Sigma_{IJ}$, $i\Sigma_{IJ}$) represent the $SO(5,C)$
generators $M_{IJ}=(L_{IJ},K_{IJ})$ in the space of the first
kind spinors.

With respect to maximal compact subgroup $SO(5)$, the Hermitian
second-rank spin-tensor $X$ may be decomposed in the complete set of
matrices $\Sigma_0$, $\Sigma_I$ and $\Sigma_{IJ}$ with the real
coefficients: 
\[
X=\frac{1}{2}\Bigg(x_0\Sigma_0+x_I\Sigma_I+
\frac{1}{2}\,x_{IJ}\Sigma_{IJ}\Bigg)\,, 
\]
and thus $(X, X)=x_0^2+x_I^2-\frac{1}{2}x_{IJ}^2\,.$ There is one more
independent invariant combination of $x_0$, $x_I$ and $x_{IJ}$
originating
from the invariant $\mbox{tr}(X\bar X)^2$ which is equivalent to
$\mbox{det}X$.

Relative to the embedding $SO(5,C)\supset SO(5)$ one has the following
decomposition in the irreducible representations: 
\[
\label{eq:16} \underline {16}=\underline 1\oplus \underline
5\oplus\underline{10}\,. 
\]
Under the discrete transformations one can get: 
\begin{eqnarray}  \label{eq:PT}
P&:&x_0\to x_0,\ x_I\to x_I,\ x_{IJ}\to -x_{IJ}\,,  \nonumber \\
T&:&x_0\to -x_0,\ x_I\to -x_I,\ x_{IJ}\to x_{IJ}\,.  \nonumber
\end{eqnarray}
From the point of view of the rotational subgroup $SO(5)$, extra times
$x_I$ constitutes the axial vector, whereas the spatial coordinates
$x_{IJ}$ constitutes the pseudo-tensor.

The rank of the algebra of $Sp(4,C)$ is $l=2$. Hence an arbitrary
irreducible representation of the noncompact group $Sp(4,C)$ is
uniquely characterized by two complex Casimir operators $I_2$ and
$I_4$ of the second
and the forth order, respectively, i.e.\ by four real quantum numbers.

Otherwise, an irreducible representation of $Sp(4,C)$ can be described
by the mixed spin-tensor $\Psi_{A_1\dots}^{\bar B_1 \dots}$ of a
proper rank. This spin-tensor should be traceless in any pair of the
indices of the same
kind, and its symmetry in each kind of indices should correspond to a
two-row Young scheme. Therefore, an irreducible representation of
$Sp(4,C)$ may unambiguously be characterized by a set of four integers
($r_1,r_2;\bar r_1,\bar r_2$), $r_1\ge r_2\ge 0$ and $\bar r_1\ge \bar
r_2\ge 0$.

The rank of the maximal compact subgroup $SO(5)\simeq Sp(4)/Z_2$ is
equal to $l=2$. Hence a state in a representation is additionally
characterized by
two additive quantum numbers, namely, the eigenvalues of the mutually
commuting momentum components of $L_{IJ}$ in two different planes,
say, $L_{12}$ and $L_{45}$.

\paragraph*{Momentum space kinematics}

The spin-tensor of the particle momentum looks like: 
\[
\Pi=\frac{1}{2}\,\Bigg(p_0\Sigma_0+p_I\Sigma_I+\frac{1}{2}\,
P_{IJ}\Sigma_{IJ}\Bigg)\,. 
\]
There are two independent invariants built of $\Pi$: 
\begin{eqnarray}
C_{{1}}&\equiv& \mbox{tr}\,\Pi\bar \Pi
=p_0^2+{p_{I}}^{2}-\frac{1}{2}\,{P_{IJ}}^2\,,  \nonumber \\
C_2&\equiv&\mbox{tr}\,(\Pi\bar \Pi)^2=\frac{1}{4} (p_0^2
+{p_{I}}^2)^2+ p_0^2\,{p_I}^{2}+{\cal O}(P^2)\,, \ \ \  P\to
0\,.
\nonumber
\end{eqnarray}
The solution to the above constraints determines the dispersion law
for a massive particle: 
\begin{eqnarray}
p_0^2&=&m^2+K\,,  \nonumber \\
{p_I}^{2}&=& C_1-m^2-K+\frac{1}{2}\,{P_{IJ}}^{2}\,.  \nonumber
\end{eqnarray}
Here 
\[
m^2=\frac{1}{2}\Big[C_{1}\pm\Big(2C_1^2-C_2\Big)^{1/2}\Big]
\]
is the rest energy squared, $K(m^2,C_1,P, \hat p)$ is the kinetic
term, $\hat p$ is the unit orientation vector of the time-momentum.
One has $K ={\cal O}(P^2)$ at $P\to 0$. Reality of $m$ and $K$
proves to require $0\le m^2\le C_1$ which results in the subsequent
restriction (cf.\ fig.~1)
\[
C_1^2 \le C_2< 2 C_1^2\,. 
\]

\begin{figure}[htbp]
\begin{center}
{\epsfxsize=70mm \epsfbox[0 0 300 215]{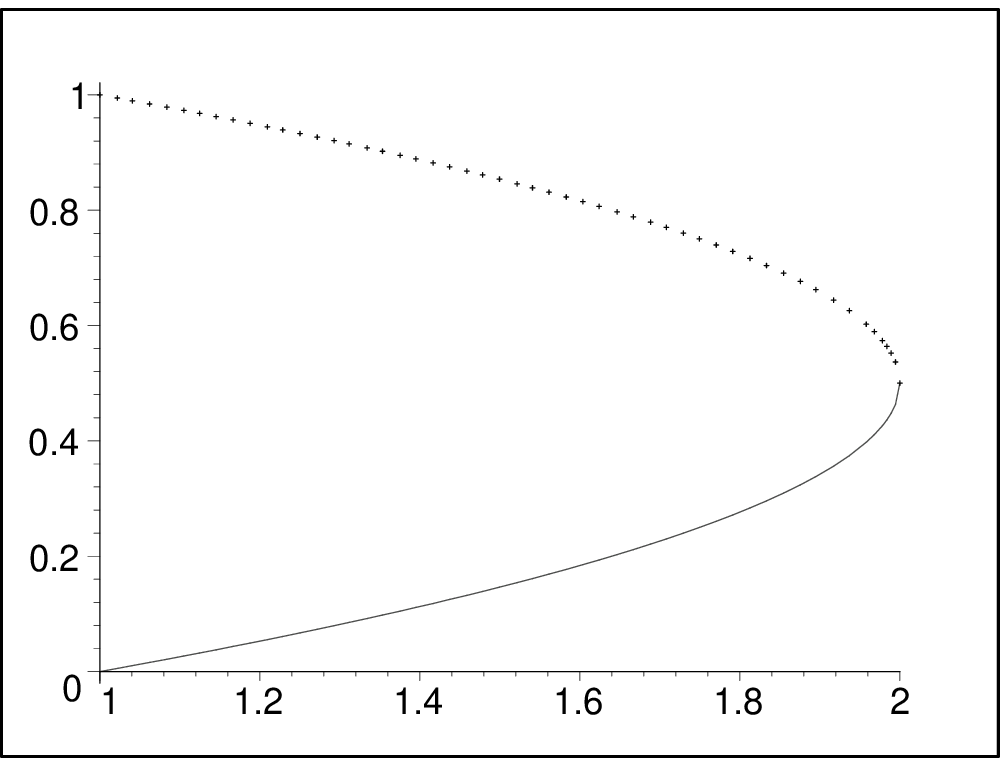}}
\end{center}
\par
\paragraph{Fig.~1}
\par
Two-valued plot of the  normalized  mass squared $m^2/C_1$ vs
normalized second invariant $C_2/C_1^2$.
\end{figure}

\paragraph*{Reduction: l $\to$ 1}

The ultimate unit of dimensionality in the symplectic approach is the
discrete number $l=1,2,\dots$ corresponding to the dimensionality $2l$
of the spinor space. The dimensionality $d=4l^2$ of the space-time
emerges as a secondary quantity. The extended space-time with $l>1$
should eventually compactify by means of the symplectic gravity via
discrete (quantum) transitions: 
\begin{eqnarray}
& \dots & \nonumber \\
&\downarrow&  \nonumber \\
& l=2&  \nonumber \\
&\downarrow&  \nonumber \\
& l=1&  \nonumber \\
&\downarrow&  \nonumber \\
& \phantom{\,.}l=0\,.&  \nonumber
\end{eqnarray}
The case $l=0$ corresponds to the (hypothetical) annihilation of the
world into ``nothing''.

For $l=2$, depending on the spinor decomposition relative to the
embedding $Sp(4,C)\supset Sp(2,C)$, three generic inequivalent types
of the space-time decomposition are conceivable: 
\begin{eqnarray}
\underline {16}&=&4\cdot \underline 4\,,  \nonumber \\
\underline {16}&=&2\cdot \underline 4\oplus \big(\underline 3+
\mbox{h.c.}\big) \oplus 2\cdot \underline 1\,,  \nonumber \\
\underline {16}&=& \underline 4\oplus \big(2\cdot\underline 2+
\mbox{h.c.}\big) \oplus 4\cdot \underline 1\,.  \nonumber
\end{eqnarray}
Note that due to compactification, the last case can result in
the violation of the spin-statistics connection in four space-time
dimensions.

\section{Summary}

To summarize: the hypothesis that the symplectic structure of
space-time is superior to the metric one provides, in particular, the
rationale for the four-dimensionality and the $1+3$ signature of the
ordinary space-time.

When looking for the space-times with extra dimensions, the hypothesis
predicts the one-parametric discrete series of the metric space-times
of the peculiar dimensionalities and signatures, with the spatial and
time extra dimensions in a definite proportion. Because one of the
time directions remains rotationally invariant under fixed boosts,
there emerges the (non-relativistic) causality despite the presence of
extra times. 

The symplectic
approach provides an unorthodox alternative to the pseudo-orthogonal
space-times and inspires a lot of new opportunities for the physics of
extra dimensions. But beyond the physical adequacy of the extended
space-times as such, by gene\-ralizing from the basic symplectic case
$l=1$ to its counterpart for general $l>1$, a deeper insight into the
nature of the very four-dimensional space-time may be attained.

\end{document}